\documentclass[12pt]{article}

\def\A{\mathsf{A}}

\def\F{\mathsf{F}}
\def\K{\mathsf{K}}

\def\SO{\mathsf{SO}}
\def\so{\mathsf{so}}
\def\SL{\mathsf{SL}}
\def\sl{\mathsf{sl}}
\def\g{\mathsf{g}}
\def\e{\mathsf{e}}
\def\U{\mathsf{U}}\def\x{\mathsf{x}}
\def\v{\mathsf{v}}
\def\f{\mathsf{f}}
\def\h{\mathsf{h}}

\usepackage{graphicx}
\begin{document}

\title{%
Doubly Special Relativity as a Limit of  Gravity}
\author{K.\ Imi\l{}kowska\thanks{email: kaim@ift.uni.wroc.pl} \ and  J.\ Kowalski--Glikman\thanks{e-mail
address jurekk@ift.uni.wroc.pl}\\  Institute for Theoretical
Physics\\ University of Wroc\l{}aw\\ Pl.\ Maxa Borna 9\\
Pl--50-204 Wroc\l{}aw, Poland} \maketitle

\begin{abstract}
Doubly Special Relativity (DSR) is a theory with two observer-independent scales, of velocity and mass, which is expected to replace Special Relativity at ultra-high energies.  In these notes we first discuss the postulates of DSR, and then turn to presenting arguments supporting the hypothesis that DSR can be regarded as a flat space, semiclassical limit of gravity. The notes are based on the talk presented at the conference ``Special Relativity -- Will it Survive the Next 100 Years?''
\end{abstract}
\maketitle

\section{Introduction}

The anniversary of a great idea is usually a good occasion for critical reassessment of its meaning, influence, and future. In theoretical physics, where methodology, instead of hermeneutics, is based on  Popperian conjectures and refutations scheme this last issue -- the future -- is, of course, the most important. Thus, in the course of the celebrations of the 100 anniversary of the Theory of Relativity, we are mostly interested in asking the questions: Is Special Relativity still to be regarded as the correct theory describing relativistic phenomena (particles and fields kinematics and dynamics) in flat space-time? Will it survive the next 100 years, and if not, which theory is going to replace it?

One quite often hears the opinion that there is, in fact, no such theory as Special Relativity. What we have to do with is just a very particular, flat space-time limit of General Relativity. And given the fact that few of us doubt that the ultimate theory of gravity should be Quantum Gravity (in the form of Loop Quantum Gravity, or String Theory, or perhaps -- and most likely -- in the disguise we  do not really know yet) the question to be posed is: what is the flat space, semiclassical limit of Quantum Gravity?

For a long time it was taken as obvious that such a limit should be just the ordinary Special Relativity. Recent developments, however, put some doubts on this naive conclusion. First, assuming that one or another form of (super)strings theory is indeed the correct theory of Quantum Gravity one may contemplate the idea that in the (super)string vacuum, corresponding to our universe, Lorentz invariance is spontaneously broken. This would certainly lead to some, possibly observable, modifications of Special Relativity \cite{bluhm}. Violation of Lorentz symmetry is also possible in models based on Loop Quantum Gravity \cite{urrutia}. The issue of Lorentz Invariance Violation and its possible observational consequences has been recently reviewed in \cite{Jacobson:2004rj} and \cite{Jacobson:2005bg}.

The core of all the proposals of spontaneous Lorentz symmetry breaking is introduction of some sort of aether. While phenomenological models of this form are certainly interesting, making possible to device precise experimental tests, they are much less appealing theoretically, since they are based on rejection of the most cherished principle of physics, the relativity principle\footnote{One should note that the presence of aether does not necessarily mean breaking of Lorentz symmetry. It is well possible that the relations between inertial observers still form Lorentz (or Poincar\'e) group, however, group elements would, in the presence of the aether, depend both on relative velocity of observers {\em and} velocity with respect to the aether. Physically, the breaking of Lorentz symmetry means that some physical processes (particle reactions, for example) are possible for some range of velocities relative to the aether and impossible for other velocities. In DSR proposal, which we will describe below, the central postulate is that relativity principle still holds, and thus if some process is observed by one observer it is observed by all other. There is also some specific relation between descriptions of the same process by two observers, which depends {\em only} on their relative velocity.}. In our view there is no a priori reason, neither theoretical, nor experimental, to contemplate  violation of relativity principle. This does not mean that modifications of Special Relativity are not possible, and we will argue below that there are good reasons to believe that if one regards Special Relativity (understood as a kinematical theory of particles and fields on flat semiclassical space-time) as a limit of (quantum) gravity, it is inevitable that some, Planck scale corrections, must be present. In a sense, the presence of this scale reflects the ``memory'' of flat space-time about its (quantum) gravitational origin.

\section{Postulates of Doubly Special Relativity}

Let us start with the plausible assumption that the flat space-time kinematics of point particles originates in some theory of quantum gravity. Assume further that in taking the flat space-time, semiclassical limit there is no breaking of spacetime symmetries, and in particular that the relativity principle holds (in other words, the diffeomorphism invariance of the theory is not being broken in the process.) After taking this limit the space-time will be, at least locally, the standard Minkowski space-time, however it may well be that the resulting theory still possesses some information about its origin, in the form of the observer independent mass scale, $\kappa$, of order of Planck scale. This scale will be still present as a parameter  in the transformation laws, relating different inertial observers.

If such a scale indeed is present,  it is natural to expect that deviations from the standard Special Relativistic kinematics   arise in the processes characterized by the energy scale close to $\kappa$, and that these deviations should be rather generic \cite{Smolin:2005cz}. For example, it may happen that the standard dispersion relation for particles acquires additional terms, to wit
\begin{equation}\label{1}
    E^2=p^2 + m^2+\alpha \frac{ E^3}\kappa + \ldots
\end{equation}
with $\alpha$ being a dimensionless parameter of order 1.

Naively, modified dispersion relation would immediately imply the existence of preferred frame, since if it holds in one frame it does not hold in any other, related to the original one by the standard Lorentz transformation. However we can demand that along with deforming dispersion relation we deform Lorentz transformations, so that (\ref{1}) (and their generalizations to be introduced shortly) remain invariant under action of six parameter group of transformations, with generators satisfying Lorentz $SO(3,1)$ algebra.

Based on this intuition, Giovanni Amelino-Camelia \cite{Amelino-Camelia:2000ge}, \cite{Amelino-Camelia:2000mn} formulated a set of postulates that should be satisfied by  new theory, replacing Special Relativity, in the regime of ultra-high energies, and he dubbed this theory Doubly Special Relativity.
\newline

Doubly Special Relativity is based on the following postulates:

\begin{enumerate}
\item {\em The Relativity Principle holds}. This means that if two observers describe the same phenomenon, possible differences in their descriptions can only depend on their relative motion (in the case of inertial observers -- their relative velocity). In particular there is no notion of absolute rest, and absolute motion. It follows also that if some process is observed by one observers (for example particle (1) collides with particle (2) producing particle (3)), all observers agree that this process takes place.
\item {\em There exist two observer-independent scales}: the velocity scale c, identified with velocity of light, and, the mass scale $\kappa$, identified with Planck mass ($\kappa\sim 10^{19}$ GeV).
\end{enumerate}

It is worth recalling at this point what is the difference between observer independent scales and other dimensionful quantities one encounters in physical theories. To be more specific, let us consider the question what is the difference, in the framework of the standard Special Relativity, between a coupling constant, like electric charge $e$ or Planck's constant $\hbar$, and the velocity of light $c$. In the first case,  of dimensionful coupling constants, all observers measure their values with the help of  identical, low energy experiments, performed in their rest frame. Then the relativity principle\footnote{In the ``passive'' form: Identical experiments performed by inertial observers give the same results.} guarantees that all the observers will obtain the same numerical values of the constants. These values can be then used in other experiments. In the case of the speed of light the situation is different however. Since this speed is an observer independent scale, we demand also that if two inertial observers measure velocity of {\em the same} photon, they will obtain the same result. This is clearly incompatible with Galilean Relativity, but, as we know, has been successfully built into the principles of his Special Relativity. Analogously, in the construction of Doubly Special Relativity we postulate the presence of yet another observer-independent scale, this time of dimension of mass, whose numerical value is presumed to be of order of Planck mass\footnote{Note, however, that  contrary to Special Relativity, in which we know physical objects moving with the velocity of light (massless particles), in the case of DSR we do not know (yet?) physical objects exemplifying the scale $\kappa$.}.

It has soon been realized \cite{jkgminl}, \cite{rbgacjkg} (for recent review see \cite{Kowalski-Glikman:2004qa})
that these postulates are satisfied in the framework of theories in which Poincar\'e algebra is replaced by deformed $\kappa$-Poincar\'e algebra \cite{Lukierski:1991pn}, \cite{Lukierski:1991ff}, \cite{kappaM1}, \cite{kappaM2}\footnote{ It should be stressed however that DSR is not just the $\kappa$-Poincar\'e algebra -- not only in the sense analogous to the well known fact Special Relativity is not just the Poincar\'e algebra -- there might be DSR proposals in which this algebra does not play any role.} .

One possible realization is provided by the, so-called, bicrossproduct (or Majid--Ruegg)basis \cite{kappaM1}, in which the commutators between rotation $M_i$, boost $N_i$, and momentum $P_\mu = (P_0, P_i)$ generators are the following 

$$
[M_i, M_j] =  \epsilon_{ijk} M_k, \quad [M_i, N_j] =  \epsilon_{ijk} N_k,
$$
\begin{equation}\label{2}
  [N_i, N_j] = - \epsilon_{ijk} M_k.
\end{equation}

and
   
$$
[M_i, P_j] =  \epsilon_{ijk} P_k, \quad [M_i, P_0] =0
$$
$$
   \left[N_{i}, {P}_{j}\right] =   \delta_{ij}
 \left( {1\over 2} \left(
 1 -e^{-2{P_0}/\kappa}
\right) + {{\mathbf{P}^2}\over 2\kappa}  \right) -  \frac1\kappa\, P_{i}P_{j},\nonumber
$$
\begin{equation}\label{3}
  \left[N_{i},P_0\right] =  P_{i}.\nonumber
\end{equation}

As one can easily check, the Casimir of the $\kappa$-Poincar\'e algebra  (\ref{2}), (\ref{3}) reads
\begin{equation}\label{4}
{\cal C} = \kappa^2\,  \cosh\, \frac{P_0}{\kappa} - \frac{\vec{P}{}^2}2\, e^{P_0/\kappa} - M^2.
\end{equation}
It is easy to check also that the expansion of (\ref{4}) to the leading order in $1/\kappa$ yields (\ref{1}).

One should note at this point that the bicrossproduct algebra above is not the
only possible realization of DSR. For example, in \cite{Magueijo:2001cr},
\cite{Magueijo:2002am} Magueijo and Smolin proposed and
carefully analyzed another DSR proposal, called sometimes DSR2.\index{DSR2} 
In DSR2 the Lorentz algebra is still not deformed and there are no deformations in the brackets of rotations and momenta. The boosts-- momenta generators have now the form
\begin{equation}\label{ms1}
   \left[N_{i}, p_{j}\right] =  i\left( \delta_{ij}p_0 -
  {1\over \kappa} p_{i}p_{j} \right),
\end{equation}
and
\begin{equation}\label{ms2}
  \left[N_{i},p_{0}\right] = i\, \left( 1 - {p_0\over \kappa}\right)\,p_{i}.
\end{equation}
It is easy to check that the Casimir for this algebra has the form
\begin{equation}\label{ms3}
 M^2 = \frac{p_{0}^2 - \vec{p}{}^2}{\left(1- \frac{p_0}\kappa\right)^2}.
\end{equation}

In order to describe kinematics of a particle we must extend the above algebra to the algebra of phase space of the particle. We will not derive here all the results, and the reader could find the derivation with references to the original literature in the review paper \cite{Kowalski-Glikman:2004qa}. Instead we will just state the main results which will be important below, when we compare DSR with a theory resulting as a flat limit of gravity coupled to point particles.

\begin{itemize}
\item[$\bullet$] Both the bicrossproduct (\ref{2})--(\ref{4}) and Magueijo--Smolin algebras (\ref{ms1})--(\ref{ms3}) can be understood as examples of larger class of algebras constructed as follows. In the standard Special Relativity four-momentum can be thought of as a point of four dimensional flat manifold of Lorentz Signature -- the flat momentum manifold. Notice that in this case positions, being ``translations of momenta'' are points of another flat Minkowski space. In transition to DSR, assume instead that the space of momenta is a maximally symmetric manifold of (constant) curvature, $1/\kappa^2$. In the limit when $\kappa$ goes to infinity, the curvature goes to zero and we return to Special Relativity, as we should. In given coordinates on this constant curvature momentum space, each point will correspond to some four-momentum.  As it is well known, the group of symmetries in this case is the 10-parameter (in 4 dimensions) de Sitter  group $\SO(4,1)$. This group possesses a six-parameter subgroup, $\SO(3,1)$, isomorphic with Lorentz group, and one can easily compute what will be infinitesimal action of the group elements on points of the manifold. It turns out that, for example, the bicrossproduct basis corresponds to the standard system of coordinates, used in cosmology. More details can be found in \cite{Kowalski-Glikman:2004qa} and \cite{Kowalski-Glikman:2003we}.

\item[$\bullet$] What about the remaining four parameters of de Sitter group $\SO(4,1)$? It is well known from differential geometry, that while the generators of $\SO(3,1)$ act as ``rotations'' the remaining ones play the role of ``translations''. This means that it is natural to identify them with positions. It turns out to be convenient to arrange the remaining four generators $x^\mu$ so as to form the Iwasawa decomposition of the $\so(4,1)$ algebra; explicitly their commutators could be brought to the following form
\begin{equation}\label{5}
    \left[x^i,x^j\right]=0, \quad \left[x^0,x^i\right]={1\over\kappa} x^i
\end{equation}
The noncommutative space-time satisfying (\ref{5}) is called $\kappa$-Minkowski space-time.

\item[$\bullet$] It should be noted that there is a natural Hopf algebra structure associated with an algebra of symmetries of de Sitter space. This algebra turns out to be exactly the quantum $\kappa$-Poincar\'e algebra of \cite{Lukierski:1991pn}, \cite{Lukierski:1991ff}, \cite{kappaM1}, \cite{kappaM2}. For more details see \cite{Kowalski-Glikman:2004tz}.

\item[$\bullet$] It should be also noted that one can in principle construct analogous structure starting from the momentum space of the particle being anti-de Sitter space \cite{Blaut:2003wg}. Explicit models in four dimensions are not known in this case, however they play a role in 2+1 gravity coupled to a particle, as we will discuss in details below.
\end{itemize}

Given characterization of properties of single particle DSR models let us now turn to the question, what is DSR coming from.

\section{Constrained BF action for gravity}

It is usually considered rather obvious that Special Relativity, regarded as a theory of particle kinematics should emerge somehow from General Relativity coupled to point particles in a limit, in which gravitational interactions are ``switched off''. It turns out that it is surprisingly difficult to prove this claim in the framework of the standard Einstein formulation of GR. First of all ``switching off'' gravity would presumably mean going to zero with gravitational constant, but this limit is known to be pathological in GR. Moreover it is well known that coupling of GR to point particle is at least problematic, if not impossible.

Our starting point must be therefore some another (but equivalent) form of the gravity action. The convenient form has been derived recently by Freidel and Starodubtsev \cite{fs}. The starting point of that paper is the observation theory of gravity can be defined by the action containing two parts: the ``vacuum'' one, being a topological field theory with an appropriate gauge group, and the constraints, leading to the emergence of the dynamical degrees of freedom of gravity. Both parts are manifestly diffeomorphism invariant, which opens new perspectives in construction of  diffeomorphism-invariant perturbation theory for quantum gravity. From our perspective, however, the most important aspect of this theory would be that it makes it possible both to define a limit, in which local degrees of freedom of gravity are switched off and the point particle coupling. One can say that in this formulation theory of gravity has the well defined ``DSR limit.''

The construction of the Freidel--Starodubtsev theory borrows from earlier works \cite{MacDowell:1977jt}, \cite{Freidel:1999rr}, \cite{Smolin:2003qu} and is based on the $\SO(4,1)$ gauge theory. The basic dynamical variables are\footnote{Below we use {\bf BOLD} typeface to denote forms (space-time indices suppressed) and {\sf SANS SERIF} typeface  to denote Lie algebra valued fields (group indices suppressed).} $\so(4,1)$ connection one form $\mathbf{A}^{IJ}$, and the $\so(4,1)$-valued two-form $\mathbf{B}^{IJ}$. The starting point is the action principle \cite{fs}
\begin{equation}\label{bf1}
S=\int \mathbf{B}^{IJ}\wedge \mathbf{F}_{IJ}
\end{equation}
where $\mathbf{F}_{IJ}$ is the curvature of connection $\mathbf{A}^{IJ}$.
The equations of motion following from this action 
\begin{eqnarray}\label{bf2}
\mathbf{F}_{IJ}=0\nonumber\\
d_\mathbf{A}\mathbf{B}_{IJ}=0
\end{eqnarray}
where $d_\mathbf{A}$ is the covariant derivative of connection $\mathbf{A}$, tell that the connection is flat, while the $\mathbf{B}^{IJ}$ field is covariantly constant. The solutions of these equations on locally connected region ${\cal U}$ is of the form 
\begin{eqnarray}\label{bfb}
\mathbf{A}=\g^{-1}\, d\, \g\quad
\mathbf{B}= \g^{-1}\, d\f\, \g, \quad \g \in \SO(4,1), \;\; \f \in \so(4,1)
\end{eqnarray}

The theory is therefore almost trivial, without any local degrees of freedom.

In order to get General Relativity we must break local symmetry of the theory from $\SO(4,1)$ down to the Lorentz group $\SO(3,1)$. To this end we denote by 5 the preferred direction  in the algebra space, and add to the action the term which explicitly breaks the $\SO(4,1)$ gauge symmetry, to wit

\begin{equation}\label{bf3}
S=\int \mathbf{B}^{IJ} \wedge \mathbf{F}_{IJ}-{\alpha \over 2} \mathbf{B}^{IJ} \wedge \mathbf{B}^{KL} \epsilon _{IJKL5}
\end{equation}
Let us now decompose the algebra index $I = (i,5)$, $i=0,\ldots,3$ with $\epsilon^{ijkl}=\epsilon^{IJKL5}$ being  an invariant $\SO(3,1)$ tensor. Note that now the first equation in (\ref{bf2}) is replaced by
\begin{equation}\label{bf2a}
    \mathbf{F}_{IJ}=\alpha\, \mathbf{B}^{KL} \epsilon _{IJKL5}
\end{equation}
and is manifestly {\em not} $\SO(4,1)$ covariant.

The $\mathbf{B}$-field enters the action $S$ only  algebraically,  so we can  substitute the solution (\ref{bf2a}) back to the action (\ref{bf3}) to get
\begin{equation}\label{bf5}
S={1 \over 4\alpha}\int \mathbf{F}^{ij}\wedge \mathbf{F}^{kl}\epsilon_{ijkl}.
\end{equation}

 It is  convenient at this point to decompose
the curvature   as follows
\begin{eqnarray}\label{bf4}
\mathbf{F}^{ij}(\mathbf{A})&=&\mathbf{R}^{ij}(\mbox{\boldmath $\omega$})-{1 \over l^2}\mathbf{e}^i \wedge \mathbf{e}^j\nonumber\\
\mathbf{F}^{i5}(\mathbf{A})&=&{1 \over l}d_{\mbox{\boldmath $\omega$}} \mathbf{e}^i
\end{eqnarray}
where $\mbox{\boldmath $\omega$}^{ij}=\mathbf{A}^{ij}$ is the 4-dimensional connection one-form and $\mathbf{e}^i=e_\mu{}^i\, dx^\mu$ is a frame field which corresponding to the metric $g_{\mu\nu}=e_\mu^i \, e_{i\nu}$. $\mathbf{R}^{ij}$ is the $\so(3,1)$ curvature of connection $\mbox{\boldmath $\omega$}$, $\mathbf{R}^{ij}(\mbox{\boldmath $\omega$})=d\mbox{\boldmath $\omega$}^{ij}+\mbox{\boldmath $\omega$}^{i}_{k}\wedge \mbox{\boldmath $\omega$}^{kj}$. Notice that for dimensional reasons we had to introduce the scale $l$, of dimension of length. 
Using the equations for the curvature (\ref{bf4}), we can  rewrite the action  in terms of $\so(3,1)$ curvature:
\begin{eqnarray}\label{bf6}
S &=& {1 \over 4\alpha}\int(\mathbf{R}^{ij}(\mbox{\boldmath $\omega$})-{1\over l^2}\mathbf{e}^i\wedge \mathbf{e}^i)\wedge(\mathbf{R}^{kl}(\mbox{\boldmath $\omega$})-{1\over l^2}\mathbf{e}^k\wedge \mathbf{e}^l)\epsilon_{ijkl}\nonumber\\
&=&-{1 \over 2G}\int(\mathbf{R}^{ij}(\mbox{\boldmath $\omega$})\wedge \mathbf{e}^k\wedge \mathbf{e}^l-{\Lambda \over 6}\, \mathbf{e}^i \wedge \mathbf{e}^k \wedge \mathbf{e}^l)\epsilon_{ijkl}\nonumber\\
&+&{1 \over 4\alpha}\int \mathbf{R}^{ij}(\mbox{\boldmath $\omega$})\wedge \mathbf{R}^{kl}(\mbox{\boldmath $\omega$})\epsilon_{ijkl}
\end{eqnarray}
What we get is nothing but the Palatini action of General Relativity with cosmological constant plus additional term whose variation vanishes identically due to Bianchi identity. Note that the Newton's constant $G$  equals  $\alpha l^2$, while the cosmological constant $\Lambda=3/l^2$. Thus the coupling constant $\alpha = G\Lambda/3$ is dimensionless and extremely small, which makes it a perfect candidate for a parameter of (both classical and quantum) perturbative expansion. As stressed by Freidel and Starodubtsev \cite{fs}, the constrained BF theory is therefore very promising as a starting point for construction of perturbative quantum gravity, where diffeomorphism invariance is manifestly preserved at all steps of perturbative expansion.

To the initial, topological action (\ref{bf1}) we can still add the $\SO(4,1)$ cosmological term of the form
\begin{equation}\label{bf7}
-{\beta \over 2}\,\int  \mathbf{B}^{IJ} \wedge \mathbf{B}_{IJ}.
\end{equation}
This addition changes the equations of motion:
\begin{eqnarray}\label{bf8}
\mathbf{F}_{IJ}-\beta\,  \mathbf{B}_{IJ}=0\nonumber\\
d_\mathbf{A} \mathbf{B}_{IJ}=0
\end{eqnarray}
(the second equation follows in fact from the first and Bianchi identity.) 

It can be shown that the action (\ref{bf7}) is still topological, i.e., without local degrees of freedom. As before
we can add to this action the $\alpha$ constraint, in order to obtain the action of General Relativity with the additional term 
$\frac2\gamma\, \mathbf{R}^{ij}(\mbox{\boldmath $\omega$})\wedge \mathbf{e}_i wedge \mathbf{e}_j$ and more more topological terms. The ``bare action'' parameters $l,\alpha,\beta$ are related to the physical ones $G$, $\Lambda$, and $\gamma$ (Immirzi parameter) as follows $\Lambda = 3/l^2$, $\gamma=\beta/\alpha$, $G=\frac{\alpha^2-\beta^2}\alpha\, l$ (for more details and discussion of possible physical relevance of $\gamma$ parameter see \cite{fs}, and also recent paper \cite{Alexander:2005vb}.)

The convenient basis of $\so(4,1)$ algebra is provided by Dirac matrices $\gamma^{ij} = \frac12 [\gamma^i, \gamma^j]$ and $\gamma^i\gamma^5$. Using the $\so(4,1)$ algebra valued fields $\sf{A}_\mu$, $\sf{B}_{\mu\nu}$ the constrained BF action for gravity can be rewritten in the following form
\begin{eqnarray}\label{bf8}
        S&=&\int d^4 x \epsilon^{\mu\nu\rho\sigma}Tr(\sf{B}_{\mu\nu}\sf{F}_{\rho\sigma}(\sf{A}))\nonumber\\
               &-& \frac\beta2\int d^4 x \epsilon^{\mu\nu\rho\sigma}Tr(\sf{B}_{\mu\nu}\sf{B}_{\rho\sigma})\nonumber\\
&-&\frac\alpha2 \int d^4 x \epsilon^{\mu\nu\rho\sigma}Tr(\sf{B}_{\mu\nu}\sf{B}_{\rho\sigma}\gamma^5)
\end{eqnarray}

It is quite easy to couple point particles to the constrained BF action. Indeed since $\mathbf{A}^{IJ}$ is a one form, it couples naturally to one-dimensional objects -- the particles world-lines.  

The general procedure of coupling particles carrying non-abelian charges to Yang-Mills potential has been developed by Balachandran, Marmo, Skagerstam, and Stern (see \cite{Balachandran:1983pc} and references therein.) In the case at hands\footnote{The results presented below have been obtained in collaboration with L.\ Freidel and A.\ Starodubtsev.} the gauge group is $\SO(4,1)$. This group acts by conjugation on its algebra, and the orbits can be labelled by two numbers, corresponding to values of two Casimirs, representing mass and spin of the particle, as follows
\begin{equation}\label{bf9}
    \K = \frac12\, m{\gamma_1\gamma^5} + \frac14\, s\, {\gamma_2\gamma_3}
\end{equation}
As the second ingredient  we take  connection, gauge-transformed by an arbitrary element $\h$ of the Lorentz subgroup $\SO(3,1)$ of $\SO(4,1)$
\begin{equation}\label{bf10}
   \A_\mu{}^\h=\h^{-1} \A_\mu \h + \h^{-1} \partial_\mu \h, \quad \h = \exp\left(\frac14\, \alpha^{ab}\gamma_{ab}\right)
\end{equation}
where, as before $\so(4,1)$ connection $\A$ decomposes into $\so(3,1)$ connection $\omega$ and tetrad $e$
\begin{equation}\label{bf11}
    \A_\mu= \left(\frac{1}{2}\, e_\mu{}^a\,{\gamma_a \gamma^5} +\frac{1}{4} \omega_\mu{}^{ab}{}\, \gamma_{ab}\right)
\end{equation}

Then the action of the particle with mass $m$ and spin $s$ coupled to constrained BF gravity is defined to be
\begin{equation}\label{bf12}
    L(z,h) = Tr\, \left(\K \A_\tau{}^\h(\tau)\right)\quad S = \int\, d\tau\, L,
\end{equation}
where $\A_\tau{}^\h \equiv \A_\mu{}^\h(z(\tau))\dot z^\mu(\tau)$ is the value of gauge transformed connection (\ref{bf10}) on the particle world-line. We see therefore that the dynamics of the particle is described with the help of the charge $\K$ it carries, and the Lorentz transformation $\h$ relating the particle rest frame and the frame of (asymptotic) observer. It can be shown that variation of the action (\ref{bf12}) leads to the correct Mathiasson-Papapetrou equations describing the dynamic of spinning particle in the presence of
torsion. When the torsion is zero we recover the usual Mathiasson-Papapetrou equation, which in the case of vanishing spin reduces to the usual geodesic equation.

Having defined the coupling of the particle to gravitational field we can address the question as to what would be the effective behavior of the particle in the limiting case, when the local degrees of freedom of gravitational field are being switched off. To answer this question, one should proceed as follows: take the action being the sum of (\ref{bf8}) and (\ref{bf12}), solve the resulting equations of motion, and then take the limit $\alpha\rightarrow0$. To see which outcomes of this procedure are possible, note that that although in this limit the gravitational field will become flat in the bulk space-time, there might be some nontrivial leftover at the worldline of the particle. This contribution of the gravitational field may lead to deformation of the (otherwise free) particle action (\ref{bf12}), leading to DSR like behavior.

Unfortunately, due to the technical difficulties, the programme described above has not been  realized in practice yet. What can be done, however, is to turn to a simpler model of gravity coupled to a particle, in 2+1 dimensions. As we will see in the next section the structure of this model is very similar to the four-dimensional case with the parameters $\alpha, \beta$ equal zero, i.e., in the limit we are mostly interested in. Moreover, the three-dimensional case is not purely of academic interest, as the following argument, borrowed from \cite{Freidel:2003sp} and \cite{Kowalski-Glikman:2004qa}, clearly shows.

The main idea is to construct an experimental  situation that
forces a dimensional reduction from the four dimensional to the $2+1$ dimensional theory. It
is interesting that this can be done in quantum theory, using the
uncertainty principle as an essential element of the argument.
Let us consider  free 
elementary particles in $3+1$ dimensions, whose mass are less
than $G^{-1}= \kappa$.  The motion of the particles will be
linear, at least in some classes of coordinates systems, not accelerating with respect to the natural inertial coordinates at infinity.  Let us consider the particle as described by an inertial
observer who travels perpendicular to the plane of its
motion, which we will call the $z$ direction.  From the point of
view of that observer, the particles are in an eigenstate of 
longitudinal momentum, $\hat{P}^{total}_z$, with some eigenvalue
$P_{z}$. Since the particles are in an eigenstate of
$\hat{P}^{total}_z$ their wavefunction  will be
uniform in $z$, with
 wavelength $L$ where (note that we assume here that $L$ is so large that we can trust the standard uncertainty relation; besides  this uncertainty relation is not being modified in some formulations of DSR)
\begin{equation}
L= {1 \over P_{z}^{total} }
\end{equation}

At the same time, we assume that the  uncertainties in the
transverse positions are bounded a scale $r$,
such that $ r \ll 2L $.
Then the wavefunction for the the particles has support on a
narrow cylinder of radius $r$ which extend
uniformly in the $z$ direction.
Finally, we assume that the state of the  gravitational field is
semiclassical, so that to a good approximation, within $\cal C$
the semiclassical Einstein equations hold.

Since the wavefunction is uniform in $z$, this implies that
the gravitational field seen by our observer will have a spacelike Killing
field $k^a= (\partial /\partial z)^a$.

Thus, if there are no forces other than the gravitational field, the semiclassical
 particles  must be described by an equivalent $2+1$ dimensional
problem in which the gravitational field is dimensionally reduced
along the $z$ direction so that the particles, which
are the source of the gravitational field, are replaced by 
punctures.

The dimensional reduction is governed by a length $d$, which is
the extent in $z$ that the system extends. We cannot take $d<L$
without violating the uncertainty principle. It is then convenient
to take $d=L$.  Further, since the system consists of the
particles, with no intrinsic extent, there is no other
scale associated with their extent in the $z$ direction. We can
then identify $z=0$ and $z=L$ to make an equivalent toroidal
system, and then dimensionally reduce along $z$. The relationship
between the four dimensional Newton's constant $G^{4}$ and the
three dimensional Newton's constant $G^{3}=G$ is given by
\begin{equation}
G^{3} = {G^{4} \over L} = {G^{4} P^{tot}_z \over \hbar}
\end{equation}

Thus, in the analogous $3$ dimensional system, which is equivalent
to the original system as seen
from the point of view of the boosted observer, the Newton's
constant depends on the longitudinal momentum.

Of course, in general there
will be an additional scalar field, corresponding to the
dynamical degrees of freedom of the gravitational field. However, since we are interested only in the four-dimensional limit, in which local degrees of freedom of the gravitational field are not present, all these fields will vanish this limit.

Now we note that, if there are no other particles or excited
degrees of freedom, the energy of the system  can
to a good approximation be described by the hamiltonian $H$ of the
two dimensional dimensionally reduced system. This is described
by a boundary integral, which may be taken over any circle that
encloses the particle.
But it is well known that in $3d$ gravity $H$ is bounded
from above. This may seem strange, but it is easy
to see that it has a natural four dimensional interpretation.

The bound is given by
\begin{equation}
M < {1 \over 4 G^{3} } = {L \over 4 G^{4} }
\end{equation}
where $M$ is the value of the ADM hamiltonian, $H$. But this just
implies that
\begin{equation}
L > 4G^{4}M = 2R_{Sch} \label{sch}
\end{equation}
i.e. this has to be true, otherwise the dynamics of the
gravitational field in $3+1$ dimensions would have collapsed the
system to a black hole!  Thus, we see that the total bound from
above of the energy in $2+1$ dimensions is necessary so that one
cannot violate the condition in $3+1$ dimensions that a system be
larger than its Schwarzschild radius.

Note that we also must have
\begin{equation}
M > P^{tot}_z ={
\hbar \over L}
\end{equation}
Together with (\ref{sch}) this implies $L>
l_{Planck}$, which is of course necessary if the semiclassical
argument we are giving is to hold.

Now, we have put no restriction on any components of  momentum or
position in the transverse directions.  So the system still has
symmetries in the transverse directions.  Furthermore, the argument
extends to any number of particles, so long as their relative
momenta are coplanar. Thus, we learn the following.

Let ${\cal H}^{QG}$ be the full Hilbert space of the quantum
theory of gravity, coupled to some appropriate matter fields, with
$\Lambda=0$. Let us consider a subspace of states ${\cal
H}^{weak}$ which are relevant in the low energy limit in which all
energies are small in Planck units.  We expect that this will have
a symmetry algebra which is related to the Poincar\'e algebra
${\cal P}^{4}$ in $4$ dimensions, by some possible small
deformations parameterized by $G^{4}$ and $\hbar$. Let us call
this low energy symmetry group ${\cal P}^{4}_{G}$.

Let us now   consider the subspace of ${\cal H}^{weak}$ which is
described by the system we have just constructed . It contains the
particle, and is an eigenstate of $\hat{P}^{tot}_z$ with large
$P^{tot}_z$ and vanishing  longitudinal momentum.
Let us call this subspace of Hilbert space
${\cal H}_{P_z}$.

The conditions that define this subspace break the  generators of
the (possibly modified) Poincar\'e algebra that involve the $z$
direction\footnote{Notice that if we assume that the four-dimensional rotational symmetry is neither broken, nor deformed, we can recover the whole 4d deformed Poincar\'e algebra from the 3d one.}. But they leave unbroken the symmetry in the $2+1$
dimensional transverse space. Thus, a subgroup of ${\cal
P}^{3+1}_{G}$ acts on this space, which we will call ${\cal
P}^{2+1}_{G} \subset {\cal P}^{3+1}_{G}$.

We have argued that the physics in ${\cal H}_{P_z}$ is to good
approximation described by an analogue system  of a particle
in $2+1$ gravity. However, as we will see in the next section the symmetry algebra acting there is not 
the ordinary $3$ dimensional Poincar\'e algebra, but  the
$\kappa$-Poincar\'e algebra in $3$ dimensions, with
\begin{equation}
\kappa^{-1}  = {4 G^{4} P^{tot}_z \over \hbar}
\end{equation}

Now we can note the following. Whatever ${\cal P}^{4}_{G}$ is,
it must have the following properties:

\begin{itemize}
\item[$\bullet$] It depends on $G^{4}$ and $\hbar$, so that it's  action on
{\it each} subspace ${\cal H}_{P_z}$, for each choice of $P_z$, is
the $\kappa$ deformed $3d$ Poincar\'e algebra, with $\kappa$ as
above.

\item[$\bullet$] It does not satisfy the rule that momenta and energy add, on
all states in $\cal H$, since they are not satisfied in these
subspaces.

\item[$\bullet$] Therefore, whatever $ {\cal P}^{4}_{G}$ is, it is not the
classical Poincar\'e group.
\end{itemize}

Let us therefore turn to  gravity coupled with point particle in 2+1 dimension.

\section{DSR from 2+1 dimensional gravity}

Even if not for the argument given in the preceding section, the 2+1 dimensional gravity coupled with point particles would be a perfect test ground for investigating properties of DSR theories. As it is well known this theory is topological, i.e., does not posses any local degrees of freedom, moreover its action
$$
S=\int d^3x Tr\left(\mathbf{e} \wedge \mathbf{F}(\mbox{\boldmath $\omega$})\right)
$$
resembles very closely the four dimensional action of the constrained BF theory in the DSR limit
$$
S=\int d^4x Tr\left(\mathbf{B} \wedge \mathbf{F}(\mathbf{A})\right)
$$

Investigations in 2+1-dimensional gravity have been pioneered by Starusz\-kiewicz  in 1963 \cite{staruszkiewicz63}, with interest revived by seminal papers by Deser, Jackiw and `t Hooft \cite{Deser:1983tn} and Witten 
\cite{Witten:1988hc}. Here we will follow the approach proposed by Matschull and Welling in \cite{Matschull:1997du}.

The action for (2+1) gravity reads:
\begin{equation}\label{3d1}
    S={1 \over 16 \pi G}\int_M d^3 x \epsilon^{\mu\nu\rho}Tr(\e_\mu \F_{\nu\rho})
\end{equation}
and the basic fields are dreibein $\e_\mu$ and antisymmetric spin connection $\omega_\mu$, whereas $G$ is the gravitational constant, which in (2+1) gravity has a dimension of inverse mass. The Lorentz group in 2+1 dimension $\SO(2,1)$ is isomorphic to $\SL(2,R)$ (which we will in the following denote just $\SL(2)$) and thus the field strength ${\F}_{\nu\rho}$  defined as: 
\begin{equation}\label{3d2}   {\F}_{\mu\nu}=\partial_\mu\omega_\nu-\partial_\nu\omega_\mu+[\omega_\mu,\omega_\nu]
\end{equation}
is Lie algebra $\sl(2)$-valued. It is convenient to assume that the dreibein $\e_\mu$ is also $\sl(2)$-valued, where this time the algebra is regarded as a vector space, isomorphic to the three dimensional Minkowski space. As a basis of the $\sl(2)$ algebra we take  three dimensional Dirac matrices in real (Majorana) representation and the trace in (\ref{3d1}) is just the matrix trace. The field equation  following from (\ref{3d1}) are 
\begin{equation}\label{3d3}
    \epsilon^{\mu\nu\rho}D_\nu {\e}_\rho=0, \quad
        \epsilon^{\mu\nu\rho}{\F}_{\nu \rho}=0
\end{equation}
The first equation implies that connection is torsion free, while the second assures the metric is flat. The general solutions of these  equations on a simply connected region $U \subset M$ is well known. It consists of the pair of scalar fields $(g,f)$, valued in the Lie group $\SL(2)$ and Lie algebra $\sl(2)$, respectively, such that
\begin{equation}\label{3d4}
\omega_\mu={\g}^{-1}\partial_\mu {\g}, \quad
    {\e}_\mu={\g}^{-1}\partial_\mu {\f}\, {\g}
\end{equation} 
where $\mathbf{g}\in SL(2)$ and $\mathbf{f}\in \sl(2)$. Note the similarity between this solution and the solution of the BF theory (\ref{bfb}).

Introduction of a particle  causes the spacetime to assume the shape of a cone with a particle placed at its top (see Figure 1.) 
The cone is characterized
 by the mass-dependent deficit angle $\alpha$:
\begin{equation}\label{3d5}
    \alpha=8 \pi G m
\end{equation}
where $m$ - particle mass, $G$ - gravitational constant. In what follows we will set $8 \pi G=1$, so that the allowed range of the mass is $m\in[0,\pi)$.
\begin{figure}[t]
\centering
\includegraphics[height=4cm]{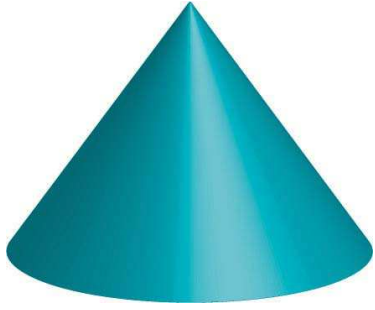}
\caption{Space-time of a single particle in 2+1 gravity has conical singularity at the top of the cone ($r=0$), being the position of the particle. The opening angle of the cone is related to the particle mass and equals $\alpha=8 \pi G m$. In the case of a spinning particle the space-time geometry is called a spinning cone characterized by deficit angle, related to the mass and time offset related to the spin, see \cite{Deser:1983tn}.}
\label{fig:1}       
\end{figure}

In polar coordinates ($t$, $r$, $\varphi$), the solution of Einstein equations corresponding to a single  spinning particle is of the form
\begin{eqnarray}\label{3d6}
    \mathbf{e}^0 &=& dt + \frac{s}{2\pi}\, d\phi\nonumber\\
    \mathbf{e}^1 &=& \left(1 - \frac{m}{2\pi}\right) \cos \phi \, dr - r \sin\phi\, d\phi \nonumber\\
\mathbf{e}^2 &=& \left(1 - \frac{m}{2\pi}\right) \sin \phi \, dr + r \cos\phi\, d\phi\nonumber\\
\mbox{\boldmath $\omega$}^0&=&\frac{m}{2\pi}\, d\phi, \quad \mbox{\boldmath $\omega$}^1=\mbox{\boldmath $\omega$}^2=0
\end{eqnarray}

This solution corresponds to particle described, similarly to (\ref{bf12}) as a delta-like singularity with an appropriate Poincar\'e charge. There is, however another, more convenient way of treating particles proposed by Matschull and Welling  \cite{Matschull:1997du}, illustrated in Figure 2. 
 As a result we get singularity free, simply connected spacetime, with boundaries. To make cylindrical boundary  look like one dimensional worldline of the particle, we take additional assumption that it circumference vanishes, which can be expressed as requirement that the component $\mathbf{e}_\varphi$ vanishes on the boundary
\begin{equation}\label{3d7}
    \bar{{\e}}_\varphi =0 \textrm{    at } r=0\textrm{, }
\end{equation}
where bar marks the value of the field on the boundary.

Since our manifold is simply connected now, a general solutions of Einstein equations in the neighborhood of the boundary  is provided by  two functions $(\mathbf {\f}(r,\varphi,t), {\g}(r,\varphi, t))$ satisfying (\ref{3d4}). On this solution we must impose appropriate boundary conditions, one of which would be (\ref{3d7}) at $r=0$, and another that guarantees continuity of dreibein and connection along the cut $r\geq0$, $\phi=0, 2\pi$. For  convenience let us denote ${\f}_\pm(r,t) = \f(r,t; \phi = 0/2\pi)$ and ${\g}_\pm =\g(r,t; \phi = 0/2\pi)$.  Since $\omega_\mu$ and ${\e}_\mu$ are to be continuous at the boundary, $\f_\pm$ and $\g_\pm$ are not independent, and are related by global Poincar\'e transformation of the form
\begin{equation}\label{3d8}
    {\g}_+={\U}^{-1}{\g}_- ,\quad {\f}_+={\U}^{-1}({\f}_- - {\v}){\U}
\end{equation}
 where ${\U} \in \SL(2)$ and ${\v} \in \sl(2)$ are constant.

\begin{figure}[t]
\centering
\includegraphics[height=4cm]{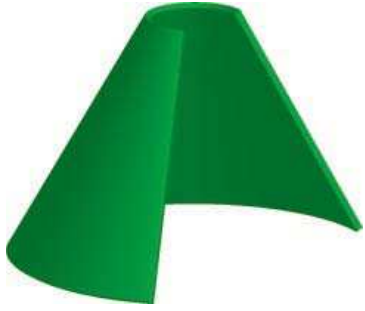}
\caption{A different way of description of a point particle has been proposed by Matschull and Welling \cite{Matschull:1997du}. Instead of a cone with singularity, they propose to use a manifold with boundaries, constructed as follows. We take the cone in Fig.\ 1, cut off its tip, and then cut the resulting surface along the line $\phi = 0 = 2\pi$. The resulting, simply connected manifold with boundaries is shown in the Figure. On this surface we introduce polar coordinates: $0 \leq r<\infty$, where the line $r=0$ corresponds to the horizontal boundary, and $0\leq \phi \leq 2\pi$, where the lines $\phi=0$ and $\phi=2\pi$ correspond to the left and right vertical boundaries. To make the surface $r=0$ looking as a worldline of a particle, we further assume that $e_\phi(r=0, t) =0$.}
\label{fig:2}       
\end{figure}

Now the condition (\ref{3d7}) along with (\ref{3d4})
 \begin{equation}\label{3d9}
    \bar{\e}_\varphi=\bar{\g}^{-1}\partial_\varphi \bar{\f}\bar{\g}=0
\end{equation}
tells that $\bar{{\f}}=\bar{{\f}}(t)$. This and the fact that boundary represents worldline the particle  makes it possible to identify $\bar{{\f}}$ with the location of the particle in space-time 
\begin{equation}\label{3d10}
\bar{{\f}}(t,\varphi)={\x}(t)   
\end{equation}
Moreover, using the condition ${\f}_+(t)=\mathbf{f}_-(t)$ and Poincar\'e transformation, we find that
\begin{equation}\label{3d11}
  {\v}={\x}(t)-{\U}{\x}(t){\U}^{-1}
\end{equation}
Taking time derivative of this equation and making use of the fact that ${\U}$ and ${\v}$ are constants gives
\begin{equation}\label{3d12}  
0=\dot{\x}(t)-{\U}\dot{\x}{\U}^{-1}
\end{equation}
This last equation is satisfied if and only if the group element $\U$ is of the form 
\begin{equation}\label{3d13}
    \U = u \mathbf{1} + p_a\gamma^a, \quad p_a\gamma^a = \frac1m\, \dot\x
\end{equation}
It is natural then to identify $p_a$ with components of momentum of the particle. Note however that this momentum has an unusual property, namely it is, geometrically, a point of the three dimensional anti de Sitter space. Indeed, since $\U \in \SL(2)$, $\det\U =1$ and thus it follows from (\ref{3d13}) that
\begin{equation}\label{3d14}
    u^2+p_0^2-\vec{p}^2=1
\end{equation}
which is just a definition of anti de Sitter space. We see therefore that three dimensional gravity coupled to point particle possesses a fundamental DSR characteristics: the energy-momentum manifold is curved\footnote{Four dimensional DSR theories with energy-momentum manifolds of the form of anti de Sitter space have not been intensively investigated, though they are known to exist, see \cite{Blaut:2003wg}. It is not clear if they arise naturally as a limit of 3+1 gravity. It is also not completely clear if de Sitter energy momentum spaces, intensively investigated in the context of DSR in four dimensions, can be obtained in the 2+1 dimensional case.}.

It can be shown that instead of the standard dispersion relation the particle on shell  satisfies the deformed equation
\begin{equation}\label{3d15}
    p^ap_a-\sin^2{m}=0
\end{equation}
Of course, in the limit when $m \ll 1$ (remember that the mass scale is set equal 1) we recover the standard dispersion relation.
\newline

We know from (\ref{3d4}) that the gravitational field in the bulk is pure gauge. It follows that when particles are present the only dynamical degrees of freedom may be associated with boundaries. Therefore, if we start with the action for gravity on the manifold with boundaries, like that in Figure 2, and then perform symplectic reduction, as the result we find action defined only on the worldline on the particle. As we will see in the moment this action  differs from the free particle one: the presence of gravitational field causes deformation of the particle lagrangian, exactly in the DSR spirit.

The procedure described briefly above has been performed by Matschull and Welling \cite{Matschull:1997du} and the resulting action reads
\begin{equation}\label{3d16}
L=-\frac{1}{2}Tr({\U}^{-1}\dot{{\U}}\, {\x})-\varsigma(\frac{1}{2}Tr({\U})-\cos m)
\end{equation}
where $\varsigma$ is the Lagrange multiplier enforcing the mass shell constraint (\ref{3d15}). Using the expression (\ref{3d13}) and the fact that $Tr(\gamma^a\gamma^b)=2\eta^{ab}$ we can rewrite the Lagrangian in the component form as follows
\begin{equation}\label{3d17}
  L = -\left(\sqrt{p^2+1}\, \eta^{ab} + \epsilon^{abc}\,p_c - \frac{p^a\, p^b}{\sqrt{p^2+1}}\right) x_b\, \dot p_a  -\varsigma\left(p^2-\sin^2 m\right)
\end{equation}
It can be shown that, in spite of the complex, nonlinear form of the Lagrangian, the resulting equations of motion are just the standard one, to wit
\begin{equation}\label{3d18}
   \dot p_a = 0, \quad \dot x^a = \varsigma\, p^a.
\end{equation}
Let us now turn to discussion of the symmetries of the particle action
\begin{equation}\label{3d19}
   S = \int d\tau\, L
\end{equation}
It is clear from the form of the Langrangian (\ref{3d17}) that the action is invariant under standard action of Lorentz generators, so that the Lorentz transformations of both position and momentum are the same as in Special Relativity.

To find generators of Lorentz transformations, we should first derive the form of Poisson brackets, resulting from the symplectic potential in (\ref{3d17}). These brackets read
\begin{equation}\label{3d20}
    \{p_a, p_b\} =0
\end{equation}
\begin{equation}\label{3d21}
    \{p_a, x^b\} =\delta_a{}^b\, \sqrt{p^2+1} + \epsilon_a{}^{bc}\, p_c
\end{equation}
and
\begin{equation}\label{3d22}
    \{x^a, x^b\} =2 \epsilon^{ab}{}_c\, x^c
\end{equation}
Note that this last bracket tells that the positions of the particle do not commute. The exact form of this bracket  differs from $\kappa$-Minkowski  type of non-commutativity (\ref{5}), but is very closely related to it \cite{Freidel:2003sp}.

Using these bracket it is not difficult to derive the form of Noether charges $J_a$, generating Lorentz transformations through Poisson bracket. These charges have the form
\begin{equation}\label{3d23}
    J_a = \sqrt{p^2+1}\, \epsilon_{abc}\, x^b\, p^c + 2 x_{[a}\, p_{b]}\, p^b
\end{equation}
and together with conserved momenta they form the standard Poincar\'e algebra.

Note however that while the action of Lorentz generators $J_a$ on space-time variables $x^a$ is purely classical, the action of translations, generated by momenta $p_a$ is deformed, as a result of the bracket (\ref{3d21}). This means that in spite of the fact that the particle lives just in Minkowski space-time the translational invariance is lost (or being deformed). This reminds somehow the model considered by Wess \cite{Wess:2003da}, in which also the isometry group is not deformed by itself, but by its action on space-time (or, more generally -- phase space) variables.

The form of the particle Lagrangian (\ref{3d16}) suggest simple generalization to the case when the energy momentum space is more general than the $\SL(2)$ manifold considered by Matschull and Welling. Consider, for example, the case when this space has the form of de Sitter space. It follows from Iwasawa decomposition of $\SO(d,1)$ group (where $d$ is dimension of space-time and momentum space) that in this case the relevant group element has the form \cite{Kowalski-Glikman:2002ft}, \cite{Kowalski-Glikman:2004tz}
\begin{equation}\label{3d24}
    \U = \exp(p_0 t_0)\exp(p_it_i) 
\end{equation}
where the generators of the ``translational'' part of  Lie algebra $\so(d,1)$ $t_0, t_i$ satisfy the commutational relations reminding the ones of $\kappa$-Minkowski space-time
\begin{equation}\label{3d25}
   [t_0, t_i] =- t_i, \quad [t_i, t_j] =0
\end{equation}
The kinetic term of the Lagrangian reads in this case
\begin{equation}\label{3d26}
    L_k = -Tr({\U}^{-1}\dot{{\U}}\, {\x}) = -\left(x^0 - p_ix^i\right)\dot p_0 -x^i\dot p_i
\end{equation}
and is invariant under action of the Lorentz group of the form of (\ref{3}), appended by an appropriate action on position variables (see \cite{Kowalski-Glikman:2003we}, \cite{Kowalski-Glikman:2004qa} for details.) In order to get the complete lagrangian, one should add to (\ref{3d26}) the term $\varsigma{\cal C}$, where ${\cal C}$ is the Casimir (\ref{4}). It can be then checked that $\kappa$-Minkowski type of non-commutativity (\ref{5}) follows from the Lagrangian (\ref{3d26}). It is not clear, however, if this Lagrangian can be obtained from gravity directly. Work on this question is in progress.

\section{Conclusions}

{\footnotesize

"Seventy-five  thousand  generations  ago,  our  ancestors  set  this
program in motion," the second man said, "and in all that time we will  be
the first to hear the computer speak."

     "We are the ones who will hear," said Phouchg,  "the  answer  to  the
great question of Life!.."

     "The Universe!.." said Loonquawl.
     
     "And Everything!.."
     
     "Alright," said Deep Thought. "The Answer to the Great Question..."

     "Yes!.."
     
     "Of Life, the Universe and Everything..." said Deep Thought.
     
     "Yes!.."
     
     "Is..." said Deep Thought, and paused.
     
     "Yes!.."
     
     "Is..."
     
     "Yes!!!?.."
     
     "Forty-two," said Deep Thought, with infinite majesty and calm.}\footnote{Douglas Adams. The Hitch Hikers Guide to Galaxy
   Fantazy. 1990.}
\newline

The current status of DSR reminds somehow the Adams' ``forty-two'', the answer to the question, which we do not really know. To be honest, we do not have any proof yet that this answer is correct, though we hope that Pierre Auger Observatory and GLAST satellite will provide us with such a proof.  However, as we tried to argue above, there are  more and more indications that the right question is ``What is the semiclassical, flat space limit of quantum gravity?'' It is our hope that it would not require seventy-five  thousand  generations to convince ourselves that this hypothesis is correct.

\section{Acknowledgement}

These notes are based on the talk presented by one of us (JKG) at the 339th WE Haraeus-Seminar Potsdam ``Special Relativity 2005: Will it Survive the Next 100 Years?'' For JKG this work is partially supported by the KBN grant 1 P03B 01828.

\end{document}